\pdfoutput=1
\documentclass{article}
\usepackage{waspaa19}
\usepackage{amsmath,graphicx,times,url}
\usepackage{color}
\usepackage[utf8]{inputenc}
\usepackage[T1]{fontenc}
\usepackage{subcaption}
\usepackage{float}
\usepackage{multirow}

\title{Investigating kernel shapes and skip connections for deep learning-based harmonic-percussive separation}

\name{Carlos Lordelo$^{1, 2}$\thanks{This work has received funding from the European Union's Horizon 2020 research and innovation programme under the Marie Skłodowska-Curie grant agreement No. 765068. EB is supported by RAEng Research Fellowship RF/128.},
      Emmanouil Benetos$^{1}$, 
      Simon Dixon$^{1}$, 
      Sven Ahlbäck$^{2}$}
\address{$^1$ Centre for Digital Music, Queen Mary University of London, London, UK\\ 
         $^2$ Doremir Music Research AB, Stockholm, Sweden\\
}

\begin{document}

\ninept
\maketitle

\begin{sloppy}

\begin{abstract}
In this paper we propose an efficient deep learning encoder-decoder network for performing Harmonic-Percussive Source Separation (HPSS). It is shown that we are able to greatly reduce the number of model trainable parameters by using a dense arrangement of skip connections between the model layers. We also explore the utilisation of different kernel sizes for the 2D filters of the convolutional layers with the objective of allowing the network to learn the different time-frequency patterns associated with percussive and harmonic sources more efficiently. The training and evaluation of the separation has been done using the training and test sets of the MUSDB18 dataset. Results show that the proposed deep network achieves automatic learning of high-level features and maintains HPSS performance at a state-of-the-art level while reducing the number of parameters and training time. 
\end{abstract}


\section{Introduction}
\label{sec:intro}
The task of Harmonic-Percussive Source Separation (HPSS) of audio signals can be seen as a particular case of audio source separation \cite{ASS_book}. Its ideal objective is to decompose a given input signal into a sum of two component signals, one containing only the harmonic sounds of the input and the other consisting of all the percussive sounds. HPSS is a useful preprocessing tool for many applications, such as the estimation of the beat of a song by analysing only the estimated percussive signal \cite{Gkiokas12}, or the implementation of time-stretching audio effects by manipulating only the harmonic components \cite{Driedger14}.

In general, percussive sounds can be seen as unpitched sounds. They are created by simultaneously stimulating a continuum of frequencies. 
On the other hand, harmonic sounds are sounds that we perceive as pitched sounds/notes. This type of sound contains a discrete and finite set of harmonic frequencies that are stimulated simultaneously. More specifically, it is possible to say that percussive sounds are well time-localised sounds with a spread out frequency behavior while harmonic sounds are frequency-localised sounds. Therefore, when analysing a time-frequency representation as the magnitude spectrogram of the input audio signal, where the vertical axis is associated to frequency and the horizontal axis to time, percussive sounds appear as vertical lines and harmonic sounds form horizontal structures. This fact is illustrated in Figure~\ref{fig:hp-sounds}.

\begin{figure}[ht]
  \centering
  \includegraphics[width=0.46\columnwidth]{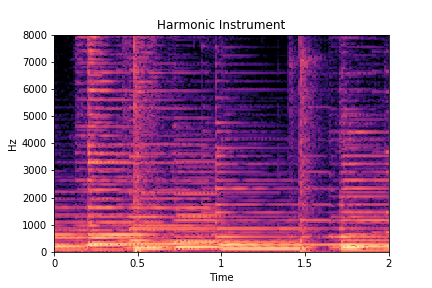}
  \includegraphics[width=0.46\columnwidth]{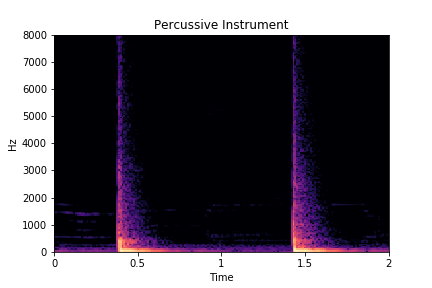}
  \caption{Real-world harmonic and percussive instrumental sounds.}
  \label{fig:hp-sounds}
\end{figure}

Traditional methods for HPSS perform the separation exploiting this particular structure of percussive and harmonic sound events in the input time-frequency representation. A good example is HPSS by median filtering \cite{Fitzgerald10}, where the segregation of the vertical and horizontal patterns is done by applying a vertical and a horizontal median filtering technique in the mixture power spectrogram. This method was later extended to an iterative process that uses different filter sizes at each iteration \cite{Driedger14_HPSS} and improved by combining source-specific proximity kernels \cite{Fitzgerald14_KAM, Liutkus14_KAM}. However, methods of this nature have intrinsic performance limitations caused by the hand-crafted filters and the strict assumptions they are based around. 

More recently, deep learning based HPSS algorithms began to rise in the literature and have shown a significant improvement over traditional HPSS methods because they can automatically learn features from data \cite{Lim17, Drossos18}. Moreover, in the Signal Separation Evaluation Campaign (SiSEC) 2018 \cite{SigSep18}, the vast majority of published methods that obtained higher performance in the task of separating music recordings into their instrumental sources (drums, vocals, bass and other) is based on Deep Neural Networks (DNNs). An architecture of particular interest is the U-Net \cite{UNet15}, where the authors proposed a deep encoder-decoder network with forward skip connections between layers of the same resolution with the objective of avoiding vanishing gradients during training and of passing directly to the decoder the high level of details of the previous feature maps. Despite being originally used for medical image processing, the U-Net has already been used successfully for singing voice separation when applied to the mixture magnitude spectrogram \cite{JanssonUNet17}. Later, it has been adapted to perform end-to-end audio separation directly in the time domain \cite{Stoller18}.         

In this paper, we propose a novel and efficient DNN for performing HPSS. The method is based on the MDenseNet \cite{MDenseNet17}, which is a variation of the U-Net with an even denser arrangement of skip connections. The MDenseNet uses fewer parameters than the traditional U-Net and achieves impressive performance when extracting a single source from the mixture \cite{MDenseNet17}. In the context of this paper, we show that it is possible to use a similar structure to perform HPSS by training the architecture in a supervised scenario to yield estimates of more than just a single source. In our case, we estimate both the harmonic and percussive sources simultaneously --- HPSS methods usually estimate only a single source and generate the other by subtracting it from the mixture. Another important contribution of our work is that we add multiple branches in our network with filters (kernels) of different shapes in the convolutional layers with the objective of allowing the network to learn the different time-frequency patterns associated with percussive and harmonic sounds more efficiently.

The remainder of this paper is organised as follows: In Section \ref{sec:methodology} we explain related works that inspired the methodology adopted by our project. Section \ref{sec:proposed_model} describes in detail the proposed multi-branch encoder-decoder network, while Section \ref{sec:data-train} details the training procedure and the dataset utilised. Then, Section \ref{sec:results} shows the results of our work and compare it with other HPSS methods. Lastly, in Section \ref{sec:conclusion} we summarise the proposed work and give an outlook of future steps.

\section{Methodology}
\label{sec:methodology}
In a DNN-based encoder-decoder architecture, it is a common practice to halve the size of the input and double the number of channels (feature maps) as the depth of network increases. The final encoded low-resolution representation is then decoded back to the original size of the input by the use of upsampling layers. This fact, however, leads to a dramatic increase in the number of parameters required by the model. 

In order to avoid this exponential explosion, we decided to adopt a different methodology in our network. We use the same strategy of the Multi-scale DenseNet (MDenseNet) \cite{MDenseNet17} to build our encoder-decoder. Instead of doubling the number of feature maps after each downsampling unit of the network, \cite{MDenseNet17} proposed to maintain constant the number of parameters at every scale of the network using DenseNets \cite{Huang2017_DenseNet} --- stacks of layers with a dense arrangement of skip connections. The authors used the MDenseNet to perform the extraction of a single instrument from a mixture and have shown that their model is able to learn feature maps more efficiently.

In here, we adapt the original MDenseNet for the task of HPSS and add extra encoder-decoder branches in our architecture with the purpose of utilising multiple kernel shapes in the internal convolutional layers. This strategy is pointed out by recent works as an efficient way to exploit the network capacity \cite{Phan16, Pons17_EfficientArch, Pons17_Timbre}. For instance, \cite{Pons17_Timbre} proposed an efficient model for timbre analysis using several filter shapes that minimised the risk of noise-fitting and over-fitting. By adopting this methodology, we propose a novel approach for the task of HPSS using a DNN with several types of convolutions, which facilitate learning harmonic--percussive relevant time-frequency patterns.

Before introducing the proposed architecture, we briefly summarise the basic blocks taken from closely related works that are used in our network, namely DenseNets or Dense Blocks \cite{Huang2017_DenseNet} and Multi-scale DenseNets \cite{MDenseNet17}. 


\subsection{Dense Block}
\label{subsec:dense-block}
In a traditional stack or block of composite layers, the output of each layer is connected just to the following one. So, we can write the output $x_{i}$ of the $i$-th layer as:
\begin{align}
x_{i} = H_{i}\{x_{i-1}\},
\end{align}
where $x_{0}$ is the original input and $H_{i}$ represents the sequence of transformations performed by the $i$-th composite layer. Observe that a stack of $L$ layers has only $L$ internal connections (counting the final output as a connection). Inspired by \cite{Huang2017_DenseNet}, we use a dense pattern of skip connections in our stack, so each layer in our block obtains additional inputs from all preceding layers and passes on its own feature-maps to all subsequent layers. Mathematically, the output of the $i$-th layer becomes:
\begin{align}
x_{i} = H_{i}\{[x_{i-1}, x_{i-2}, x_{i-3}, \dots , x_{0}]\},
\end{align}
where $[\dots]$ represents the concatenation operation. In our block, the $i$-th layer has now as input a concatenation of $i$ outputs, consisting of the feature maps of all preceding layers and the original input. This fact increases the number of internal connections of an $L$-layered block to $\frac{L(L+1)}{2}$. Such dense connectivity pattern was first introduced in \cite{Huang2017_DenseNet} and is illustrated schematically in Figure \ref{fig:densenet} for $L = 4$.

\begin{figure}[ht]
  \centering
  \centerline{\includegraphics[width=0.85\columnwidth]{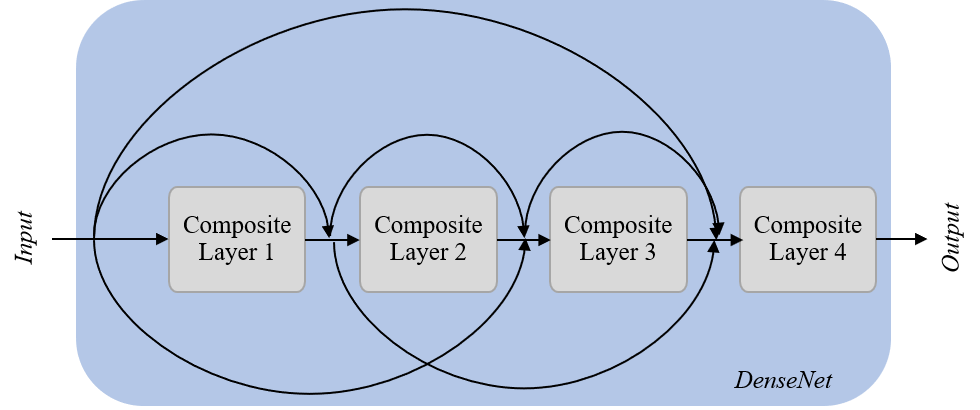}}
  \caption{Dense block with $4$ layers.}
  \label{fig:densenet}
\end{figure}

By interconnecting all layers in this way, we ensure maximum information flow between the layers of the network, which not only reduces vanishing gradients during training, but also allows the network to require fewer parameters than traditional networks, as there is no need to relearn redundant feature maps due to the re-utilisation of features already computed in the preceding layers.

In the context of our work, each composite layer consists of a convolutional layer followed by batch normalisation with a ReLu activation function. For the rest of the paper, we are considering the number of channels (output feature maps) of each convolutional layer as constant and denoted by $k$, while the number of layers in the dense block is denoted by $L$. Moreover, $k$ is referred to as growth rate since the number of input channels in each composite layer grows linearly with $k$ --- supposing the input has $n$ channels, the number of feature maps in the input of the $i$-th convolutional layer is $n + (i-1)k$. In our case, we are performing the separation using monoaural music recordings, so $n = 1$. 


\subsection{Multi-Scale DenseNet}
\label{subsec:MDenseNet}
The Multi-scale DenseNet (MDenseNet) \cite{MDenseNet17} is an encoder-decoder network whose architecture is similar to a U-Net \cite{UNet15}. However, instead of the traditional convolutional layers with exponential increase of the number of channels as the layers go deeper, MDenseNet uses dense blocks with a small growth rate. It has been shown that the MDenseNet can reduce the number of parameters while maintaining a high level of performance in extracting a single source from the mixture \cite{MDenseNet17}.

As the model goes deeper (decrease in scale), there is a loss in the resolution of the feature maps due to the addition of pooling layers (downsampling units). This fact makes the decoder unable to rebuild higher resolution feature maps with the same level of detail they had before being downsampled by the previous encoder layers. Hence, just like a U-Net, the MDenseNet also includes forward skip connections from the encoder path to the corresponding scale in the decoder path, which allows the decoder to directly receive the high level of detail of the earlier feature maps. An example of an MDenseNet architecture with depth $4$ is schematically shown in Figure \ref{fig:MDenseNet}. In the context of our work, the dense blocks of an MDenseNet have the same growth rate $k$ and the same number of layers $L$ independently of the scale at which they are located. Moreover, we utilise max-pooling layers and transposed convolutions with kernel shape of $2\times2$ as downsampling and upsampling units, respectively.

\begin{figure}[htb]
  \centering
  \centerline{\includegraphics[width=\columnwidth]{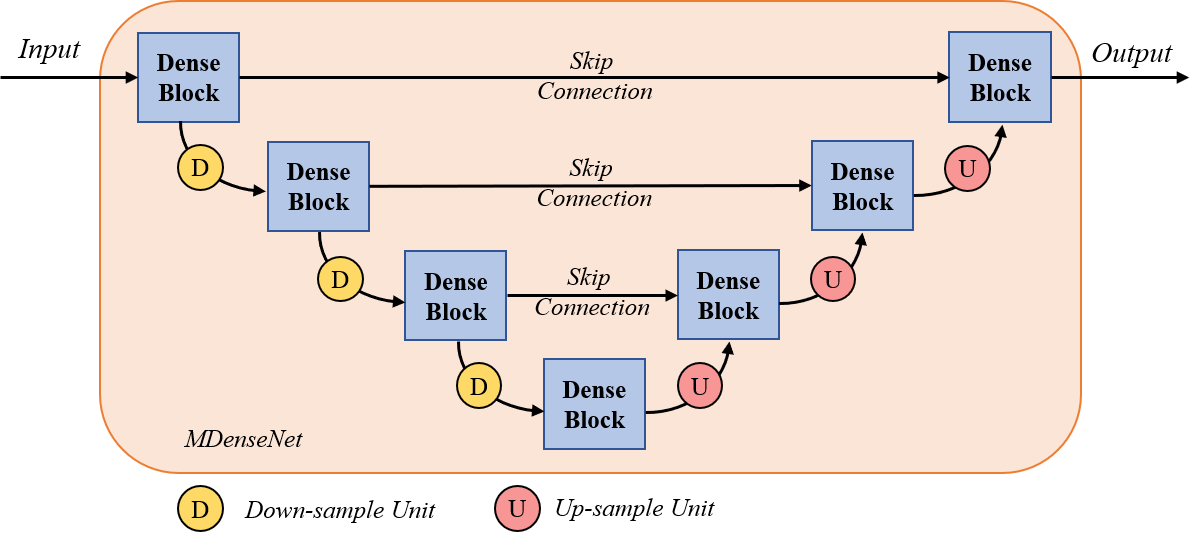}}
  \caption{Block diagram of the MDenseNet with a depth of $4$.} 
  \label{fig:MDenseNet}
\end{figure}

\section{Proposed Architecture}
\label{sec:proposed_model}
Our proposal is to adapt the original MDenseNet encoder-decoder structure by including convolutional layers with filters of different shapes to facilitate learning of the percussive and harmonic patterns present in the input spectrogram. In order to do that, we propose the \emph{Three-way Multi-scale DenseNet} ($3$W-MDenseNet), where instead of using a single encoder-decoder path, we use $3$ separate branches of MDenseNets that are combined later. Each branch uses a different filter shape for its internal convolutional layers.

We use a branch with the conventional squared kernel shape of $3\times3$ for all the convolutional layers, and we include $2$ extra branches, where all the internal convolutional layers have kernel shapes of $13\times1$ and $1\times13$ respectively. Those values were set based on the vertical and horizontal patterns that percussive and harmonic components form in the mixture spectrogram. It is expected that those new filter shapes will help the model to learn those patterns more easily and efficiently. Note that we still use a branch with square filters because no real-world musical instrument is strictly percussive or harmonic --- we still need to learn the percussive-like patterns of the transient part of harmonic instrumental sounds and associate them correctly as well as noisy-like patterns of percussive sounds. 

\begin{figure}[htb]
  \centering
  \centerline{\includegraphics[width=\columnwidth]{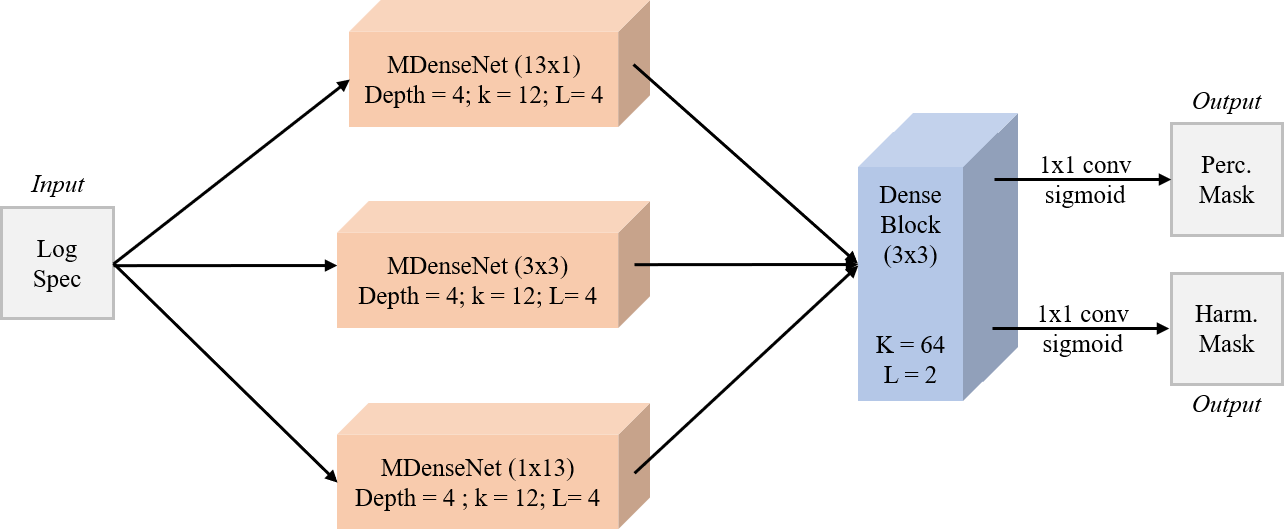}}
  \caption{Proposed architecture of the $3$W-MDenseNet.}
  \label{fig:hpss_model}
\end{figure}

The proposed architecture for HPSS is depicted in Figure \ref{fig:hpss_model} with the necessary parameters to re-implement the model. Each branch processes the same input spectrogram in parallel and their outputs are concatenated in a final Dense Block with a LeakyReLu activation function. Percussive and harmonic soft masks are computed by two separated $1\times1$ convolutional layers with sigmoid activation functions. The estimated time-domain signals are then generated by multiplying the respective masks by the input magnitude spectrogram and computing the inverse STFT using the mixture phase. All the convolutions use `valid' zero-padding to ensure the final concatenation and internal skip connections can be done correctly.

\section{Dataset and Training Details}
\label{sec:data-train}
The dataset utilised for training and testing the model is MUSDB18 \cite{musdb18}. It was the official dataset used for the evaluation of the separation models submitted to the SiSEC 2018 \cite{SigSep18}. Its corpus consists of $150$ music recordings of high quality (sampling rate of $44.1$kHz) and the associated stems for the original vocals, bass, drums and `other' source signals. MUSDB18 is organised into two directories: one containing $100$ recordings to be used as a train set and another with $50$ recordings to be used as a test set.

After performing the STFT of the monoaural time-domain audio signal using a window length of $1024$ samples with $50\%$ of overlap, an input magnitude spectrogram with size $512$x$128$, which corresponds to approximately $1.5$ seconds, is given as input to the model. Afterwards, a min-max normalisation procedure is performed over the log1p, i.e., $\log (1 + x)$, version of the input. Such normalisation is done using the global minimum and maximum value over all the possible spectrograms in the training set. 

Supervised training is done using the drums stem as ground-truth for the percussive source and the difference between the original mixture and the drums stem as ground-truth for the harmonic source. We use $80$ of the $100$ music recordings of the MUSDB18 test set for training and the other $20$ recordings as a validation set. The optimiser used is ADAM and the learning rate is reduced if the validation loss does not decrease for $3$ epochs. After the validation loss ceases improving for $15$ epochs we stop the training process.

\begin{table*}[ht]
\caption{Objective evaluation of HPSS. The values are in dB and represent the mean of the metrics over all $50$ songs in the test set.}
\centering
\begin{tabular}{|l|lll|lll|lll|}
\hline
\multicolumn{1}{|c|}{\multirow{2}{*}{Model}}       & \multicolumn{3}{c|}{Percussive} & \multicolumn{3}{c|}{Harmonic} & \multicolumn{3}{c|}{Average} \\ \cline{2-10} 
\multicolumn{1}{|c|}{}                             & \multicolumn{1}{l}{SDR}         & \multicolumn{1}{l}{SIR}       & \multicolumn{1}{l|}{SAR} & \multicolumn{1}{l}{SDR} & \multicolumn{1}{l}{SIR}         & \multicolumn{1}{l|}{SAR}      & 
\multicolumn{1}{l}{SDR}  & \multicolumn{1}{l}{SIR} & \multicolumn{1}{l|}{SAR} \\ \hline
U-Net         & $3.16$ & $4.32$ & $5.27$      & $9.56$ & $10.84$ & $12.15$     & $6.36$ & $7.58$ & $8.71$    \\ \hline
3W-U-Net      & $3.25$ & $4.34$ & \boldmath{$5.80$}      & $9.80$ & \boldmath{$10.98$} & $12.26$     & $6.53$ & $7.66$ & $9.03$    \\ \hline
MDenseNet     & $3.48$ & $4.92$ & $5.49$      & \boldmath{$9.93$} & $10.84$ & $12.46$     & \boldmath{$6.71$} & $7.88$ & $8.98$    \\ \hline
$\mathbf{3}$\textbf{W-MDenseNet}  & \boldmath{$3.70$} & \boldmath{$5.84$} & $5.35$      & $9.71$ & $10.48$ & \boldmath{$13.32$}     & \boldmath{$6.71$} & \boldmath{$8.16$} & \boldmath{$9.34$}    \\ \hline
\end{tabular}
\label{tab:results}
\end{table*}

The loss function optimised during the training is a linear combination of the mean squared error between the magnitude spectrogram of the estimated sources and the respective ground-truths. Supposing the mixture magnitude spectrogram is denoted by $X$ while the ground-truth for the percussive part is $P$ and for the harmonic part is $H$, we can define the loss function $\mathcal{L}$ as:
\begin{align}
    \mathcal{L} = \lambda_{P}||(M_{P} \odot X) - P||_{F}^2 + \lambda_{H}||(M_{H} \odot X) - H||_{F}^2,
\end{align} where $M_{P}$ and $M_{H}$ are the estimated time-frequency masks for the percussive and harmonic sources respectively, $\odot$ represents the Hadamard (element-wise) product and $||\dots||_{F}$ the Frobenius norm. Since we are interested to obtain maximum performance on the separation of both sources, we decided to set the weights $\lambda_{P}$ and $\lambda_{H}$ as $0.5$, but this value can be modified according to the application.

\section{Experimental Results}
\label{sec:results}
We implemented and evaluated $3$ additional models and compared to the proposed $3$W-MDenseNet. The first is the traditional U-Net \cite{JanssonUNet17} with squared filters of shape $3\times3$ with exponential growth at each scale. This model serves as a baseline approach and uses neither dense blocks in its architecture nor multiple filter shapes. The second is a modified version of the U-Net adopting the same design strategy of multiple branches with different kernel shapes. This model uses the same filter shapes utilised by the $3$W-MDenseNet in a similar multi-branch architecture, but does not contain any dense blocks. It is denoted by $3$W-U-Net in the rest of the paper. The third method is a version of the original MDenseNet \cite{MDenseNet17} using only squared $3\times3$ filters. All the methods use ReLu activation functions in their internal layers with a final sigmoid activation function to estimate masks for both percussive and harmonic sources.

To properly evaluate the models' capacity, we kept the number of total trainable parameters of each method in the range of $550$k--$610$k. It is important to note that the proposed $3$W-MDenseNet has around $555$k parameters, which is really lower than most of the current DNN-based methods for audio source separation. For instance, the U-Net proposed by \cite{JanssonUNet17} uses $8.7$ millions of parameters, \cite{Uhlich17} uses a BLSTM for source separation that uses more than $6$ million parameters, and \cite{Mimilakis17} proposes a complex recurrent encoder-encoder approach for singing voice extraction that uses a total of $24$ million parameters. 

We evaluate the performance of the $4$ models on the default test set of MUSDB18, which has a total of $50$ music recordings --- a considerably high number of samples if compared to just the $80$ music-recordings that were used for training. The HPSS performance is measured with the Signal-to-Distortion Ratio (SDR), Signal-to-Interference Ratio (SIR), and Signal-to-Artifact Ratio (SAR), which have been computed using the museval\footnote{https://github.com/sigsep/sigsep-mus-eval} Python package. The results are expressed in dB in Table \ref{tab:results} and the highest values for each metric appear in boldface. 

It can be seen that the proposed architecture obtained better average performance on all the metrics. The results also show that, by just adding filters of different shapes in the traditional U-Net architecture, the performance of the HPSS is already improved over all the metrics. This fact consolidates the idea that different filter shapes increase network capacity and facilitate learning of high-level features. 

\begin{figure}[htb]
  \centering
  \includegraphics[width=\columnwidth, height=5cm]{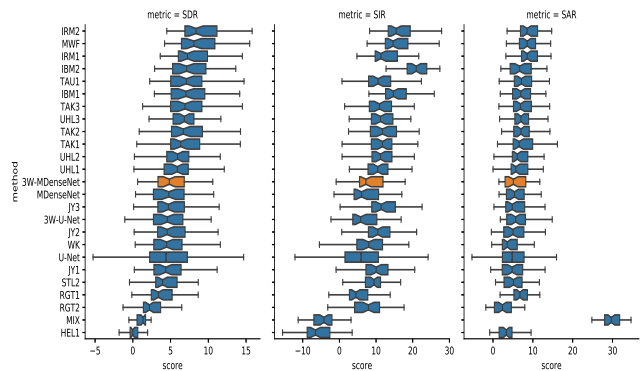}
  \caption{Comparison of the SiSEC 2018 results for percussive separation with the results obtained in our experiments.} 
  \label{fig:drum_result}
\end{figure}

Since the evaluation of SiSEC 2018 was also performed over the MUSDB18 dataset, we compared our results for the percussive separation with the results published on the campaign \cite{SigSep18} for drums extraction. The comparison is depicted in Figure \ref{fig:drum_result}, where there are box-plots sorted by the highest to the lower SDR. Comparing the $3$W-MDenseNet with the original MDenseNet, we can see that adding multiple kernel shapes reduced the variance of the overall HPSS performance and also improved considerably the SIR value --- this can also be concluded from Table \ref{tab:results}, where one can notice an increase of $0.9$~dB on the SIR value of the percussive separation.      

More information about the experiments and audio samples can be found in the experiment repository\footnote{http://c4dm.eecs.qmul.ac.uk/WASPAA19-HPSS/}.





\section{Conclusions and Future Work}
\label{sec:conclusion}
In this paper we proposed the $3$W-MDenseNet, an extension of the original MDenseNet architecture to perform HPSS. This novel network combines vertical, squared and horizontal filters in its internal convolutional layers with the objective of facilitate learning of time-frequency patterns associated with percussive and harmonic sounds. As indicated by our experiments, it outperforms both the U-Net and the MDenseNet architecture when trained under comparable settings. Our model is able to maintain a high level of performance with a low number of parameters if compared to the majority of state-of-the-art DNN-based methods. 

In this work we have not used any form of data augmentation. We believe that by adding techniques of this nature and by using other datasets, the model will achieve a better performance. It is important to remember that we trained the model only on $80$ songs and tested on $50$.  

Since our model is able to output more than a single source, we plan to investigate the possibility of adapting this model to work with other types of instrumental sources as a next step. We believe this can be achieved if we add other kernel shapes and dilated convolutions in the architecture. Due to the harmonically spaced frequency patterns of harmonic instrument sounds, dilated convolutions can potentially increase even more the network capacity.







\bibliographystyle{IEEEtran}
\bibliography{refs}
%
%
%
%
%
%
%
%
%

\end{sloppy}
\end{document}